# DCMSA: Multi-Head Self-Attention Mechanism Based on Deformable Convolution For Seismic Data Denoising

Wang Mingwei[1,2], Li Yong[1,2*], Liu Yingtian[1,2], Peng Junheng[1,2], Li Huating[1,2]

[1]  *Key Laboratory of Earth Exploration and Information Technology of Ministry of Education, Chengdu University of Technology, China.*

[2]  *State Key Lab of Oil and Gas Reservoir Geology and Exploitation; School of Geophysics, Chengdu University of Technology, China.*

## ABSTRACT

When dealing with seismic data, diffusion models often face challenges in adequately capturing local features and expressing spatial relationships. This limitation makes it difficult for diffusion models to remove noise from complex structures effectively. To tackle this issue, we propose a novel convolutional attention mechanism Multi-head Self-attention mechanism based on Deformable convolution (DCMSA) achieving efficient fusion of diffusion models with convolutional attention. The implementation of DCMSA is as follows: First, we integrate DCMSA into the UNet architecture to enhance the network's capability in recognizing and processing complex seismic data. Next, the diffusion model utilizes the UNet enhanced with DCMSA to process noisy data. The results indicate that this method addresses the shortcomings of diffusion models in capturing local features and expressing spatial relationships effectively, proving superior to traditional diffusion models and standard neural networks in noise suppression and preserving meaningful seismic data information.

# INTRODUCTION

Seismic data processing is an important part of geophysical exploration, and seismic noise attenuation is also an essential part. With the complexity of the oil and gas exploration environment, the random noise interference during seismic data acquisition has become more serious. In the past, researchers have proposed numerous methods for noise suppression, and traditional methods such as Fourier transform (Zhou et al., 2015), Wavelet transform (Sinha et al., 2005), and Curvelet Transform (Starck et al., 2002) in the domain transform method. A denoising method based on a sparse representation of data in the direction of the algorithm (Chen, 2017). Processing methods based on mathematical analysis tools such as singular value decomposition, and empirical modal decomposition (Bekara and van der Baan, 2009). Denoising methods based on seismic inversion, such as Bayesian inversion. Machine learning has received widespread attention with the updated iteration of computing devices in recent years. As a branch of machine learning, deep learning has also been widely used in geophysics. Such as the twice-denoising autoencoder framework for random seismic noise attenuation by Liao et al. (Liao et al., 2023). The framework of this article consists of a denoising autoencoder and a data generator, which can effectively remove most of the noise and retain valid information through two clever denoising. Some scholars proposed A convolutional autoencoder method for simultaneous seismic data reconstruction and denoising (Jiang et al., 2021), which can protect seismic signals while suppressing random noise and improving the resolution of seismic signals. However, these methods are poor in robustness and prone to collapse during training.

Due to the excellent robustness of the diffusion model and the ability to generate high-quality samples in the field of generative models (Croitoru et al., 2023), we chose to use the diffusion model to process seismic data. The diffusion model transforms the process of attenuation of seismic data noise into a step-by-step prediction task process by utilizing Bayesian equations. The diffusion model is divided into the forward diffusion process and the reverse diffusion process. In the forward diffusion process, clean seismic data becomes noisy data by adding noise step by step. Therefore, the noise removal process can be regarded as the inverse of the forward diffusion process, i.e., the reverse diffusion process. Ho et al. proposed the Denoising Diffusion Probabilistic Model in 2020 (Ho et al., 2020). However, due to the limited sampling range of traditional fixed convolutional kernels, the UNet in the diffusion model does not perform well in the face of feature map details during the sampling process. For these reasons, although the diffusion model has a strong sample generation ability(Peng et al., 2024), the diffusion model is not ideal for processing complex and diverse seismic data. To solve this problem, we propose the DCMSA module. We constructed a UNet network containing DCMSA and used this UNet network to process seismic data and pass the processed seismic data to the diffusion model. Then the noise is gradually predicted and removed by the inverse diffusion process of the diffusion model. Through sample experiments and evaluation analysis, we finally trained a model that can effectively reduce the noise of seismic data and retain most of the effective information.

# METHODS

The traditional attention mechanism in the diffusion model relies on neurons to extract and transfer information, but it is difficult to capture local features and spatial relationships, resulting in limited shape information transfer, while convolution can break through these limitations. Therefore, many scholars have pursued the integration of convolution with attention mechanisms (Ullah et al., 2023; Woo et al., 2018). By combining Multi-head Self-attention with Deformable convolution, we propose DCMSA, which realizes the efficient fusion of convolution and attention mechanism and improves the network's ability to sample feature maps and handle the complex structure of seismic data.

**Principle and structure of DCMSA**

DCMSA incorporates a Multi-head Self-attention mechanism with Deformable convolution to improve the network's ability to extract data features. The ability of the model to capture information is increased by incorporating Deformable convolution into Multi-head Self-attention, and the sampling position of the convolution kernel is adjusted by utilizing the offsets learned from the deformable convolution layer(Dai et al., 2017), which makes it possible to ultimately obtain a feature map with richer information. As shown in Fig. 1, figure (a) shows the traditional multi-head self-attention module and (b) shows the DCMSA module. We abandon the fully connected layers of the traditional attention mechanism and use convolutional layers instead. The convolutional layers are useful for feature map processing as they can capture local spatial information and maintain the spatial structure of the input feature map compared to the fully connected layer.

The specific process of DCMSA model input and output is shown in Fig. 1(b), as Input feature maps are deformed through the Deformable convolutional (Dconv) layers of Q(Query), K(Key), and V(Value), and the corresponding feature maps after sampling are generated by convolutional sampling of the Deformable convolutional layers. The feature map sampled by the Q, K, and V is split into multiple heads along the channel dimension, and each head corresponds to a subspace of the head dimension for attention calculation. The Q, K, and V are then reshaped using "Reshape" to satisfy the shape of the attention computation, respectively. and then computed by the "Scaled Dot-Product Attention" to obtain the normalized attention weight matrix. Then the attention weight matrix is reshaped into the same shape as the input Q, K, and V. The reshaped Q, K, and V are passed through the "Projection Convolution" layer for feature fusion to form a feature map that satisfies the output of the network. For the network to learn the optimal combination between the output of the attention mechanism and the original input and to improve the stability of network training, we then use the "Residual Connection" to connect the feature map with the original input to obtain the final feature map output.

**Principle and structure of Deformable convolution**

Compared with the traditional fixed convolutional kernel, the convolution kernel of Deformable convolutional can better handle complex and diverse data distribution, and improve the generalization of the model. By introducing learnable offsets, Deformable convolution allows the kernel to adjust its sampling points during convolution operations, as illustrated in Fig. 2. Unlike the fixed convolution kernel, which samples only at the positions of the nine blue balls, Deformable convolution can sample surrounding areas, such as the pink balls, while also sampling the nine blue balls. This capability enables the convolution kernel to adapt to geometric deformations within images, enhancing its ability to capture the shape and structural features of the target. Consequently, Deformable convolution proves advantageous for processing complex structural data.

The structure and learning process of Deformable convolution is shown in Fig. 3. Compared with the traditional fixed convolution kernel, Deformable convolution has an extra offset part. First, the obtained feature map is taken as input, a convolution layer is applied to the feature map, and then the convolved feature map is offset, which makes the convolution sampling position shift, and the offset sampling position is obtained. The offset sampling positions are then fused with the original convolution sampling positions to obtain a convolution network with both the original sampling positions and the offset sampling positions.

**Application of DCMSA to Diffusion Model**

To improve the ability of the diffusion model to process seismic data and solve the difficulty of the diffusion model in removing the noise of complex parts of the seismic structure, we added the DCMSA mechanism to improve its sampling and processing ability of complex seismic data.

*Theory and structure of the diffusion model*

Denoising diffusion theory is mainly divided into two parts: the Forward diffusion process and the Backward diffusion process.

Forward diffusion process: We add noise to clean data step by step. Let the clean seismic data be $x_0$ and the noisy seismic data be $x_t$. It is assumed that the noise increases gradually from $x_0$ to $x_t$, and that the additional noise intensity is set to the sequence β at each step, as shown in Figure 4.

So $x_t$ can be expressed as:

$$x_t = \sqrt{\alpha_t} * x_{t-1} + \sqrt{1-\alpha_t} * z_t \tag{1}$$

Among them are:

$$\alpha_t = 1 - \beta_t \tag{2}$$

$z_t$ denotes the Gaussian-distributed noise at the moment t. Also, $x_{t-1}$ can be expressed as:

$$x_{t-1} = \sqrt{\alpha_{t-1}} * x_{t-2} + \sqrt{1-\alpha_{t-1}} * z_{t-1} \tag{3}$$

Substituting equation (3) into equation (1) gives:

$$x_t = \sqrt{\alpha_{t-1} * \alpha_t} * x_{t-2} + \sqrt{(1-\alpha_{t-1}) * \alpha_t} * z_{t-1} + \sqrt{1-\alpha_t} * z_t \tag{4}$$

Where $z_t$ and $z_{t-1}$ are both random noises that obey normal distribution and are independent of each other. According to the additivity of the normal distribution, it is denoted by $z'$:

$$\sqrt{(1-\alpha_{t-1})*\alpha_t}*z_{t-1}+\sqrt{1-\alpha_t}*z_t=\sqrt{1-\alpha_{t-1}*\alpha_t}*z' \quad (5)$$

Therefore, the joint equations (4) and (5), $x_t$ can be expressed as:

$$x_t=\sqrt{\alpha_{t-1}*\alpha_t}*x_{t-2}+\sqrt{1-\alpha_{t-1}*\alpha_t}*z' \quad (6)$$

obtained in the same way, by stepwise recursion:

$$x_t=\sqrt{\bar{\alpha}_t}*x_0+\sqrt{1-\bar{\alpha}_t}*\hat{z}_t$$

where $\bar{\alpha}_t$ denotes the cumulative multiplication of $\alpha_1$ to $\alpha_t$ and $\hat{z}_t$ denotes the combination of $z_1$ to $z_t$ Gaussian distributions. Therefore, throughout the orthogonal process, we derive the formula from $x_0$ to $x_t$ which realizes the orthogonal modeling with stepwise noise addition.

Backward diffusion process: in the Backward diffusion process, we invert the solution $x_{t-1}$ by Bayesian equation, which can be expressed by using the conditional probability formula as:

$$P(x_{t-1}|x_t,x_0)=P(x_t|x_{t-1},x_0)*\frac{P(x_{t-1},x_0)}{P(x_t,x_0)} \quad (8)$$

where the right-hand side of equation (8) can be expressed as:

$$P(x_t|x_{t-1},x_0)=\sqrt{\alpha_t}*x_{t-1}+\sqrt{1-\alpha_t}*z_t \quad (9)$$

$$P(x_{t-1},x_0)=\sqrt{\bar{\alpha}_t}*x_0+\sqrt{1-\bar{\alpha}_t}*\hat{z}_t \quad (10)$$

$$P(x_t,x_0)=\sqrt{\bar{\alpha}_t}*x_0+\sqrt{1-\bar{\alpha}_t}*\hat{z}_t \quad (11)$$

Following other related work (Dhariwal and Nichol, 2021; Song et al., 2020b), we then use the Gaussian distribution function:

$$x\sim N(\mu,\sigma^2) \quad (12)$$

$$f(x)=\frac{1}{\sqrt{2*\pi}*\sigma}*\exp\left(-\frac{(x-\mu)^2}{2*\sigma^2}\right) \quad (13)$$

Expanding equation (9) gives:

$$x_{t-1} \sim N(\mu_{t-1}, \sigma_{t-1}^2) \tag{14}$$

$$\mu_{t-1} = \frac{1}{\sqrt{\alpha_t}} * (x_t - \frac{1-\alpha_t}{\sqrt{1-\overline{\alpha}_t}} * \hat{z}_t) \tag{15}$$

$$\sigma_{t-1}^2 = \frac{(1-\overline{\alpha}_t) * \beta_t}{1-\overline{\alpha}_t} \tag{16}$$

So far, throughout the process of backward diffusion, we have derived $x_{t-1}$ in terms of $x_t$. Similarly, $x_{t-2}$ can be calculated in the same way in terms of $x_{t-1}$, which makes it possible to calculate $x_0$ step by step. At the same time, this theory has been proven to be valid by many studies in many fields (Saharia et al., 2022b).

*DCMSA-based UNet network*

The UNet network is the core module of the entire diffusion model network. UNet uses a typical encoder-decoder structure. The encoder performs successive convolution and downsampling to generate a feature map with a small resolution. Then, the decoder continuously convolves and upsamples to the original size to obtain segmentation results. In the diffusion model, the UNet network mainly helps the diffusion model process the input feature map.

The Attention Gate proposed by Oktay et al. is effective in highlighting salient features useful for a specific task (Oktay et al., 2018). The Deformable Large Kernel Attention proposed by Azad et al. benefits from Deformable convolution, which allows for flexible distortion of the sampling network, enabling the model to adapt to different data patterns (Azad et al., 2024). Sun et al. proposed a Residual Encoder based on a simple attention module to improve the extraction of fine-grained features by the backbone. The semantic representation of a given feature mapping is reconstructed through multi-head self-attentive attention to the lowest-level features, further enabling fine-grained segmentation of different classes of pixels (Sun et al., 2022). Inspired by these scholars' articles, to improve the ability of diffusion models to better represent features and model more details, we have built our DCMSA model in the UNet architecture so that the network can enhance the modeling ability of local details and learn richer and high-quality feature representations. The specific structure is explained below.

In this paper, we add a module containing DCMSA, which we call D-Block, to the UNet network, and the overall structure of UNet is shown in Fig. 5. The D-Block in the UNet network in the figure is responsible for sampling the previous feature maps, extracting the image features and transferring them to the next step, which is composed of a Deformable convolution, a DCMSA module, a convolution, and a Residual Module. The flow of the figure is to first sample the input feature map through deformable convolution and modify the number of data channels, then process the feature map through the DCMSA module, then return the number of channels through ordinary convolution, and finally through the residual, so that the output data has the new features of the sampled data while retaining the features of the original data.

In summary, the UNet network in this paper adds a module containing DCMSA to the basic UNet network structure (Ronneberger et al., 2015), which first trains the ability to predict noise through the diffusion model, and then uses the UNet network to process the feature map and pass it to the diffusion model, and then denoises. Finally, good results were achieved. As shown in Figure 6, both samples show good denoising results.

EXPERIMENT

**Experimental equipment and parameters**

In this paper, the model was trained and validated using an NVIDIA RTX 3060 Ti GPU with 8 GB of memory. It is beneficial for the subsequent interpretation of seismic data, and the experimental hyperparameters of the specific Diffusion model and UNet are shown in Table 1.

In this experiment, the first comparison method we used was the Diffusion model proposed by researchers such as Durall et al. (Durall et al., 2023). The second comparison method is TDAE by Liao et al.

**Data set description**

Synthetic data were used for model training in this experiment, which included a variety of geological formations involved in actual exploration tasks. After training, the synthetic data are used for performance evaluation, and then the field seismic data are processed and validated. To increase the generalizability of the network, we choose data of different complexity and features as much as possible when training the network. The synthetic data used in this case are SEAM and Marmousi models. Figure 7 illustrates the 6 types of samples of the synthetic data.

**Experimentation and Evaluation**

*Synthetic data validation*

In Figure 8, we select some synthetic seismic data to be trained and compared on the Diffusion model, TDAE, and DCMSA. In the figure, we can see that although all three methods have noise residuals, DCMSA retains most of the effective information while removing most of the noise, the residual noise of the Diffusion model and TDAE is more obvious, and the loss of effective information is more serious.

We take the data in Fig. 8 as an example, and show the noise removal images of the three methods of TDAE, Diffusion model, and DCMSA in Fig. 9, i.e., the resultant graphs of the noise difference between the data before denoising and the data after denoising. From the whole figure, it can be seen that the noise density removed by DCMSA and TDAE is higher compared to the Diffusion model, which indicates that the denoising ability of DCMSA and TDAE is more excellent. In the place with complex structures, as shown in Figure 9, the position of the red box is most obvious, especially the position circled by the red wireframe of Data3, and TDAE removes the relatively high noise density at the position of the red box of Data3, which corresponds to the complex structure in the red box of the noise data Data3, indicating that TDAE has excellent ability to identify and process complex structures. In Data3, the Diffusion model does not have a higher noise density in the red box, and its noise density is similar to that of the rest of the whole figure, indicating that its ability to identify complex structures and denoising noise is weak. If we look at Data3 of DCMSA, the noise density of DCMSA in the red box is significantly higher than that of TDAE and Diffusion model, and it is consistent with the complex structure of noise data, indicating that DCMSA has better identification and denoising processing of complex structures than TDAE and Diffusion model. The reason for DCMSA's better ability to deal with complex structures, we analyze is that DCMSA samples more details, which makes DCMSA recognize the complex structure part during training, and makes the network learn the complex structure features of the data effectively so that it has more excellent ability to better performance in dealing with complex structures.

To be able to compare the differences between the three methods in Fig. 9 more clearly, we introduced MSE to analyze their denoised data, and the MSE formula is as follows:

$$MSE = \frac{1}{m}\sum_{i=1}^{m}(y_i - \hat{y}_i)^2$$

Where $(y_i - \hat{y}_i)^2$ denotes the difference between the denoised data and the clean data. MSE determines the superiority of the denoising effect by comparing the mean square error between the denoised data and the clean data, i.e., the smaller the MSE is, it means that the smaller the gap between the denoised data and the clean data, and the better the denoising effect is. Figure 10 shows the MSE histogram analysis of the three methods of DCMSA, Diffusion model, and TDAE, in which it can be seen that in the MSE of the three groups of samples arranged from small to large, DCMSA is the smallest, followed by TDAE, and lastly, Diffusion model. It indicates that the DCMSA processed data is most similar to the clean data, and also reflects that DCMSA has the best denoising effect among the three methods and retains the most effective information. Since the two samples (a) and (b) in Fig. 9 are more complicated, the MSE values of the two methods are higher, but the MSE value of DCMSA is still smaller than that of the Diffusion model and TDAE, which indicates that DCMSA is also more superior in dealing with the complex geological structures above.

*Methodological Evaluation*

Signal-to-Noise Ratio(SNR): The SNR is a commonly used metric in signal processing to measure the ratio of useful information to noise in a signal. In this paper, we choose the average SNR under different noises as the evaluation index. In this part of the experiment, we first add noise to the synthetic validation dataset to generate noisy data. We generated four levels of randomized noise data. The average SNR of the three methods is shown in Fig. 11. With the enhancement of the SNR of the red folded noise data, the SNR of the three methods is improving, but the SNR of the denoised data of the DCMSA, whether at low SNR or high SNR of the noisy data, have different degrees of improvement compared with the Diffusion model and the TDAE. It shows that DCMSA has a strong adaptive ability in dealing with different noise levels.

Structural Similarity (SSIM): SSIM is a metric used to measure the degree of similarity between two images, and is commonly used in the fields of image processing, computer vision, and image quality evaluation. The value of SSIM often ranges from -1 to 1. The closer the value is to 1, the more similar the two data are. Among the three methods, we chose the same number of training times 200 times, validated 50 sets of data respectively, obtained the denoised data, and then calculated the SSIM between the clean data and the denoised data, if its SSIM is closer to 1, it reflects the better denoising effect. The average SSIM of each method is shown in Table 2. From the data in the table, it can be seen that the SSIM value of DCMSA is larger than other methods, which indicates that its preservation of the data structure is more complete and the denoising effect is more excellent.

*Real Data Experiment*

To verify the effectiveness of this method on complex actual data and highlight the advantages of DCMSA, we selected the seismic data of a construction area in the South China Sea for testing, as shown in Fig. 12, we selected three real data samples to verify our method. We validate the advantages of DCMSA by comparing the Diffusion modeling method using the DCMSA module with the Diffusion model without the DCMSA module on real data.

From Fig. 12, we can see that DCMSA denoises better in the complex structure part of the graph, and the Diffusion model loses more effective information in the complex structure part. To make the difference between the two methods more obvious, we use the sample (a) of Figure 12 as an example to construct the noise map of the real data removal in Figure 13. When comparing the Diffusion model with the DCMSA, the DCMSA removes significantly more noise than the Diffusion model. In detail, there are partial faults (circled in red boxes) in the raw data plot of Figure 13. In the Diffusion model diagram, the noise density of the Diffusion model is similar to that of other parts of the complex structure, indicating that it is not targeted enough for the complex structure part. However, compared with DCMSA, DCMSA has a significantly higher noise density in the tomographic part (the red box part of the DCMSA diagram), which indicates that through the detailed sampling of DCMSA, the network has targeted learning and denoising in the complex structure part of the data, so that in the face of complex structure, DCMSA removes more noise and retains more effective information.

# CONCLUSION

In this paper, A Multi-head Self-attention mechanism based on Deformable convolution (DCMSA) is proposed and applied to the denoising task of seismic data. We deal with the noise of seismic data by establishing a diffusion model framework that includes DCMSA. Compared to traditional diffusion models, DCMSA is better at handling the overall structure and complexity of seismic data. We conducted experiments and evaluation analysis on synthetic and real data, and the results showed that our proposed method performed well in both data scenarios, effectively identifying and dealing with more noise while reducing signal leakage. Therefore, we believe that the method proposed in this paper has significant potential and significance for improving the accuracy of seismic exploration and reconstructing weak signals.

# LIST OF FIGURES

LIST OF TABLES

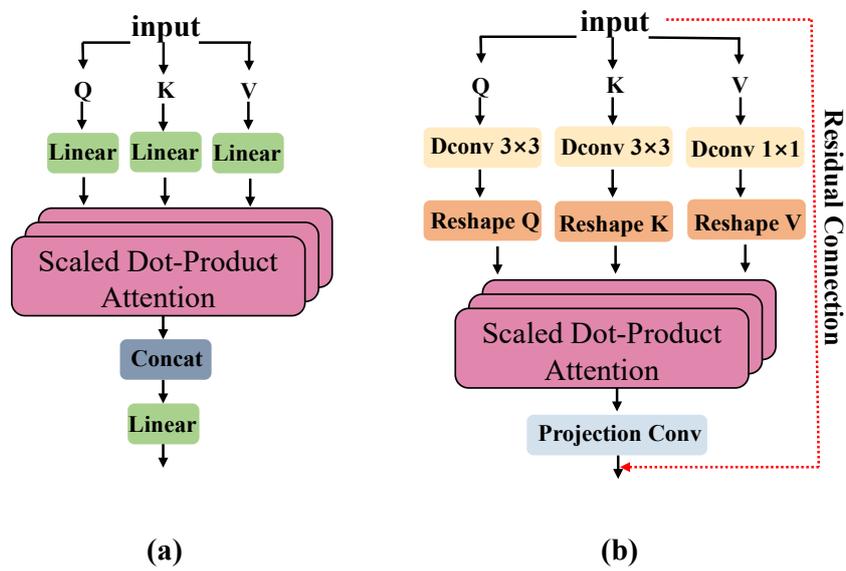

Figure 1. (a) Structure of the Multi-head Self-attention mechanism; (b) Structure of the DCMSA mechanism

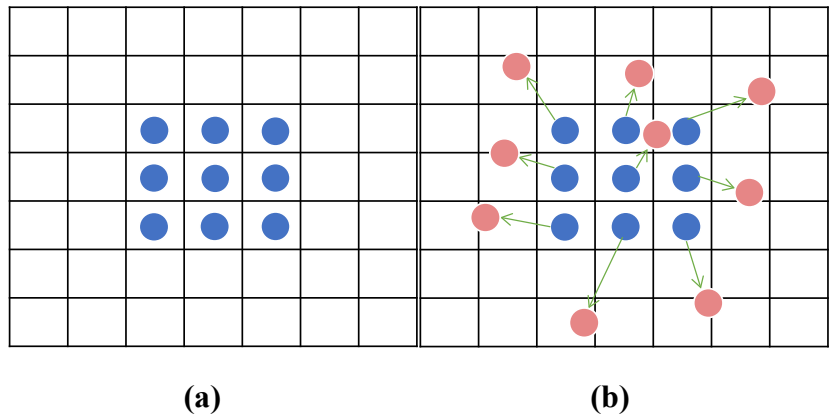

14. Fig 2. Comparison of the ordinary convolutional kernel and Deformable convolutional kernel, (a) the ordinary 3×3 convolution kernel, (b) 3×3 Deformable convolutional kernel (where the blue ball represents the fixed sampling position and the pink sphere represents the dynamic sampling position)

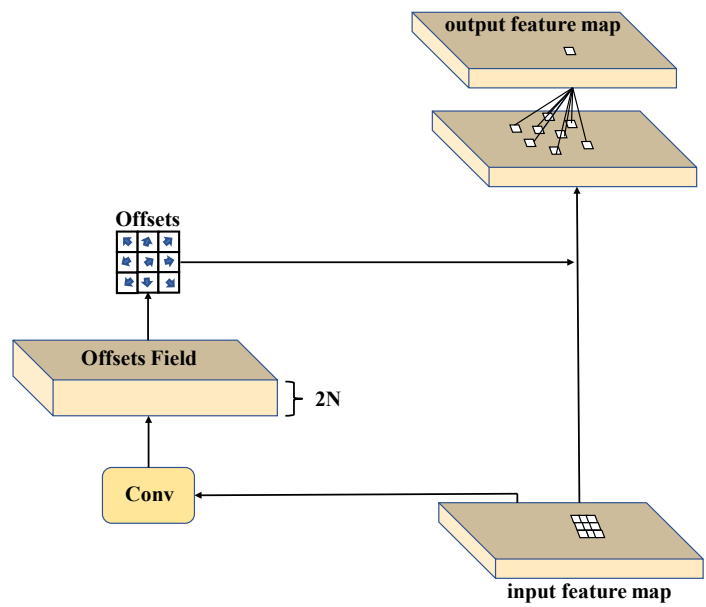

Fig 3. Structure of Deformable convolution

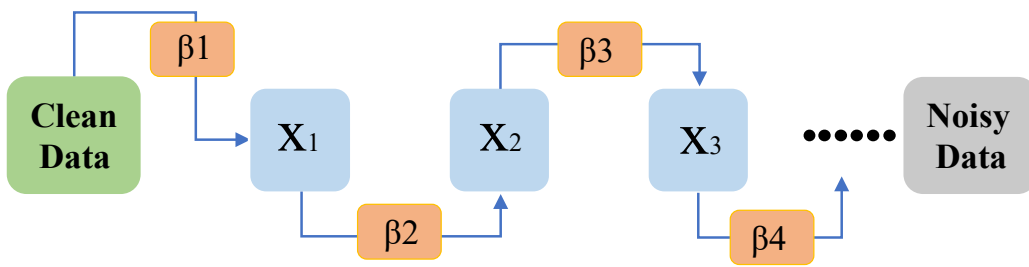

Figure 4. The process of adding noise to the forward diffusion process

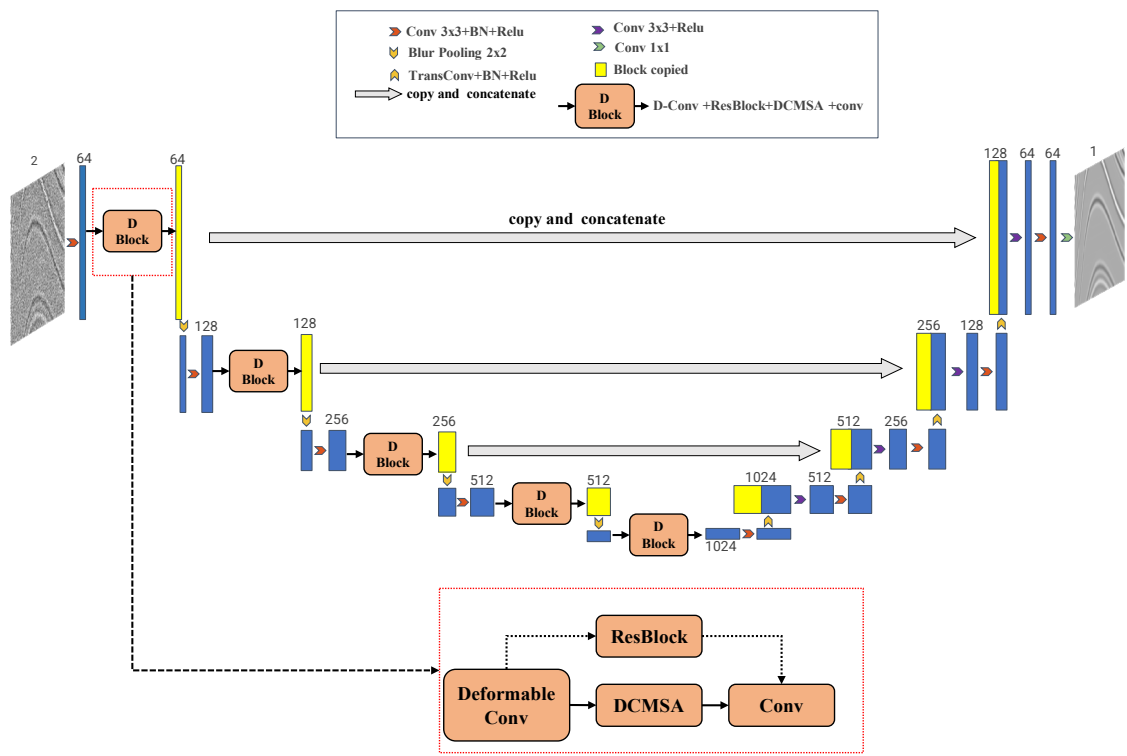

Fig.5 Structure of DCMSA-UNet network

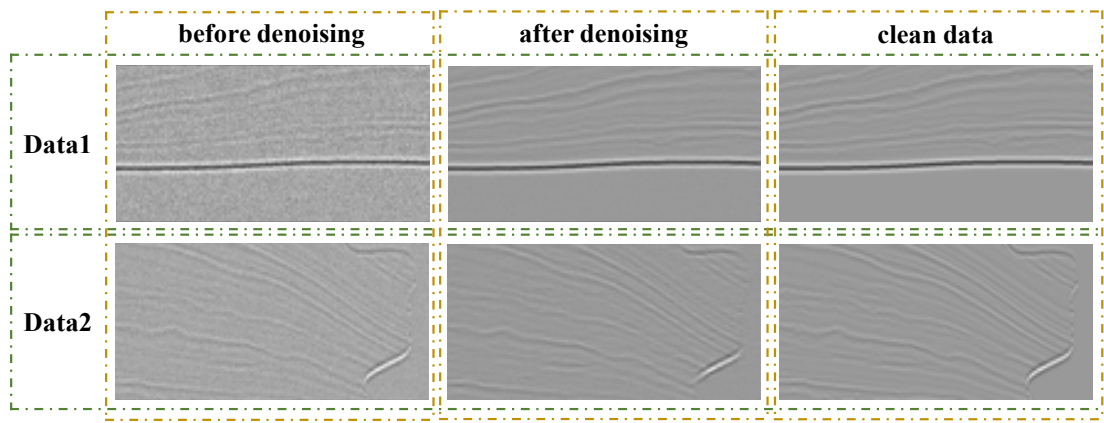

Figure 6 The denoising effect of this method on synthetic data

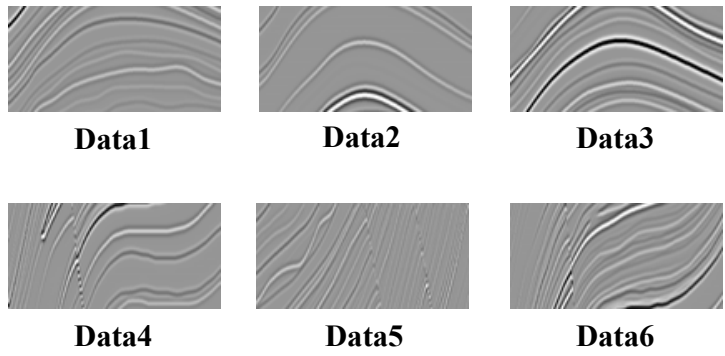

Figure 7. Example of a dataset

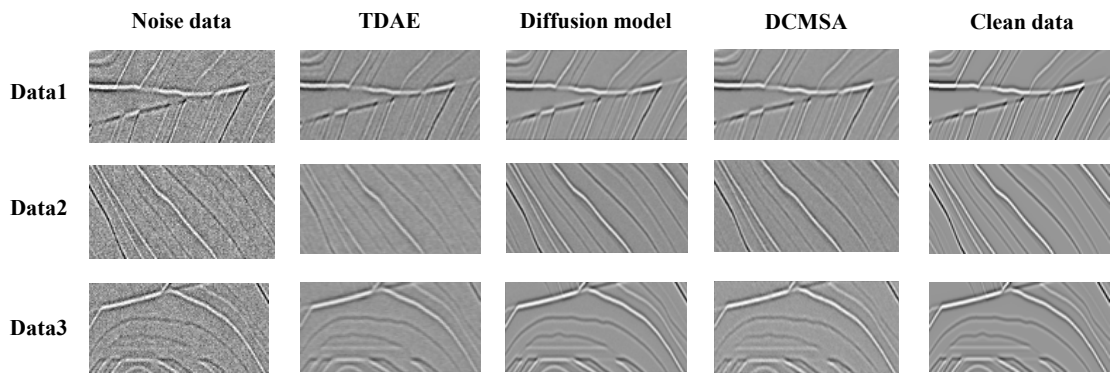

Figure 8. Method comparison chart

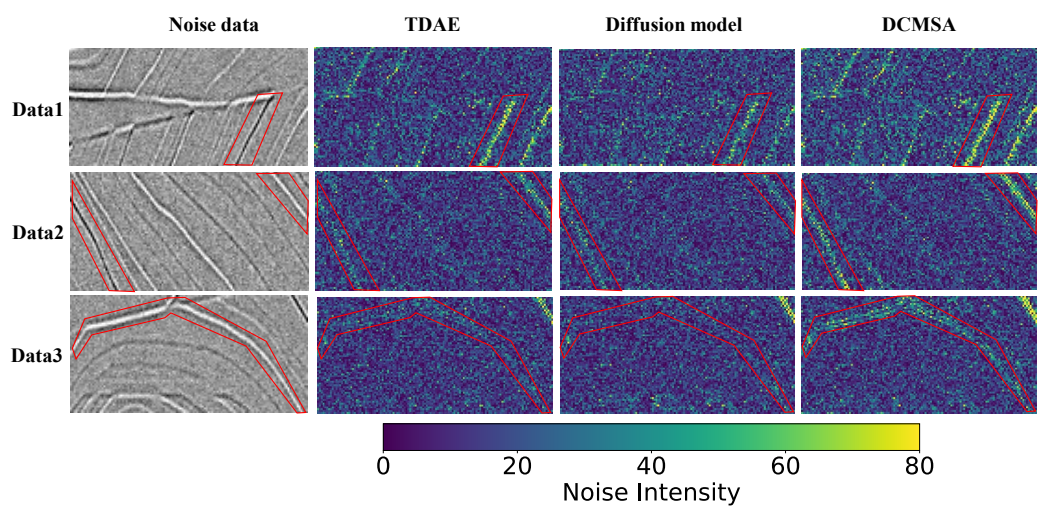

Fig 9. Synthetic data noise removal maps

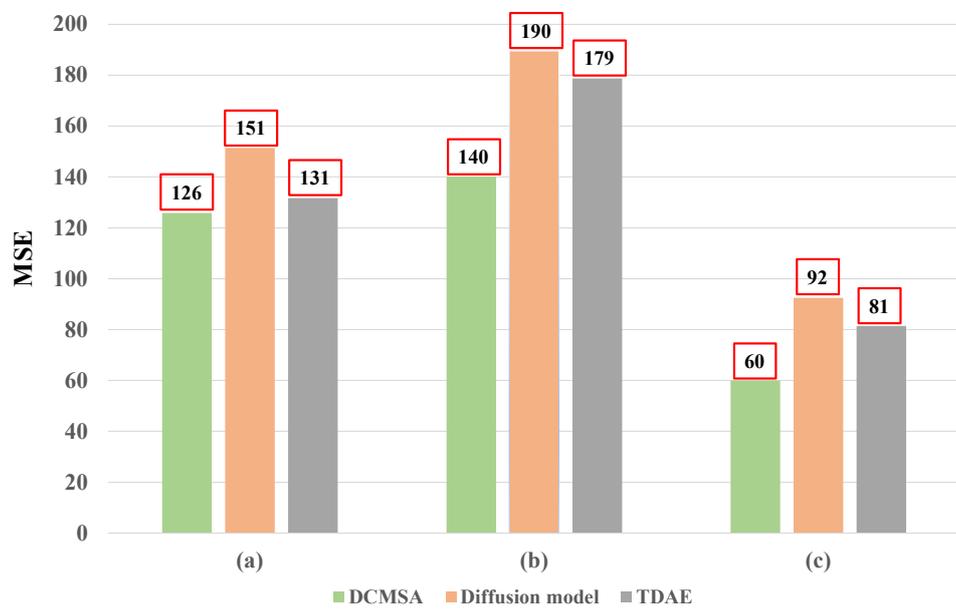

Fig 10. MSE method analysis

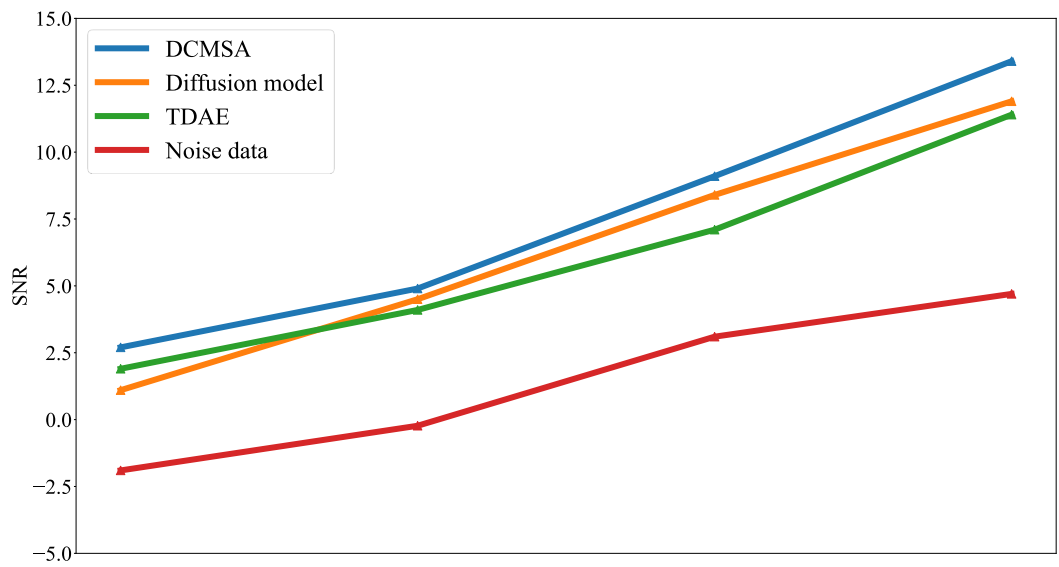

Fig.11 Average SNR of denoising data

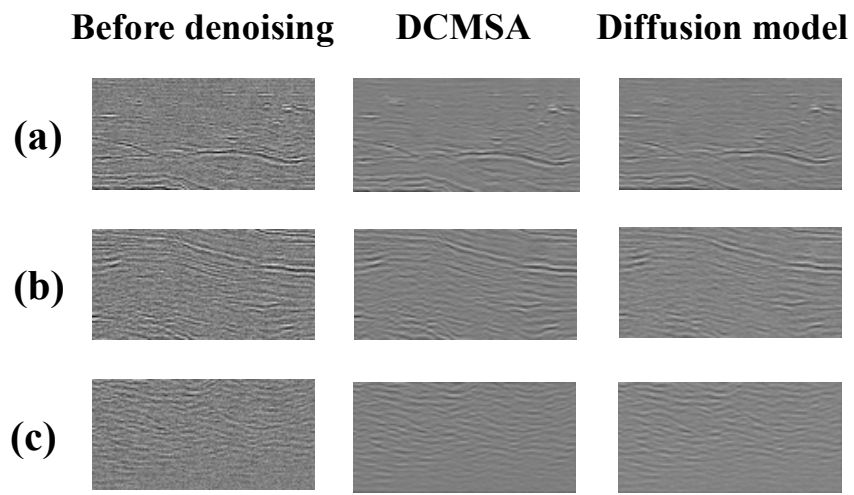

Figure 12. Comparison of denoising effects on real data

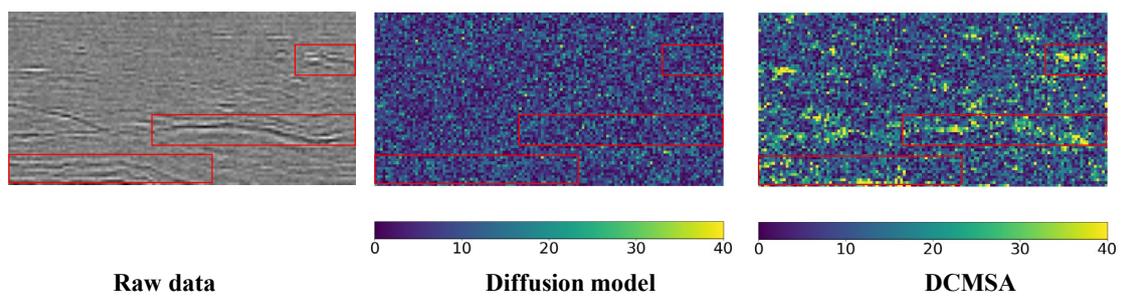

Fig 13. Noise map for real data removal

| Diffusion model | | UNet | |
|---|---|---|---|
| batch size | 32 | batch size | 16 |
| learning rate | 0.000003 | learning rate | 0.00002 |
| time step | 2000 | periodicity | 1000 |
| loss function | L1 | gradient | 2 |
| EMA | 0.999 | EMA | 0.995 |

Table 1 Training hyperparameters

| method | value |
|---|---|
| TDAE | 0.722 |
| Diffusion model | 0.770 |
| DCMSA | 0.854 |

Table 2 Comparison of Average SSIM